# Evaluating a Healthcare Data Warehouse For Cancer Diseases

Dr. Osama E.Sheta
Department of Mathematics (Computer Science)
Faculty of Science, Zagazig University
Zagazig, Elsharkia, Egypt
oesheta75@gmail.com

Ahmed Nour Eldeen
Department of Mathematics (Computer Science)
Faculty of Science, Zagazig University
Zagazig, Elsharkia, Egypt
ahmednour_cs@yahoo.com

*Abstract*— **This paper presents the evaluation of the architecture of healthcare data warehouse specific to cancer diseases. This data warehouse containing relevant cancer medical information and patient data. The data warehouse provides the source for all current and historical health data to help executive manager and doctors to improve the decision making process for cancer patients. The evaluation model based on Bill Inmon's definition of data warehouse is proposed to evaluate the Cancer data warehouse.**

## I. INTRODUCTION

Managing data in healthcare organizations has become a challenge as a result of healthcare managers having considerable differences in objectives, concerns, priorities and constraints. The planning, management and delivery of healthcare services included the manipulation of large amounts of health data and the corresponding technologies have become increasingly embedded in all aspects of healthcare. Information is one of the most factors to an organization success that executive managers or physicians would need to base their decisions on, during decision making. Healthcare organizations typically deal with large volumes of data containing valuable information about patients, procedures, treatments and etc. These data are stored in operational databases that are not useful for decision makers. The concept of "data warehousing" arose in mid 1980s with the intention to support huge information analysis and management reporting [1]. The data warehousing provides a powerful solution for data integration and information access problems. Data warehousing idea is based on the online analytical processing (OLAP). Basically, this technology supports reorganization, integration and analysis of data that enable users to access information quickly and accurately [2]. Data warehouse was defined According to Bill Inmon a "subject-oriented, integrated, time variant and non-volatile collection of data in support of management's decision making process" [3]. According to Ralph Kimball "a data warehouse is a system that extracts, cleans, conforms, and delivers source data into a dimensional data store and then supports and implements querying and analysis for the purpose of decision making"[4]. Evaluation is the final stage in the development of data warehouse where the cleansed and finalized data are evaluated against some acceptance criteria, such as uniqueness, applicability, representative, provability, validity, understand-ability etc.

## II. RELATED WORK

Several approaches have been adopted to evaluate a data warehouse. Some of them look at a broader view where the entire data warehousing project is assessed; others focus in a specific view where only the data warehouse is examined.

Approaches that evaluate the entire data warehousing project, such as the one proposed by David Heise [5], consider the following elements as listed in Table 1 below:

TABLE 1: ELEMENTS TO EVALUATE A DATA WAREHOUSE

| # | Element | Description |
|---|---|---|
| 1 | Methodology | Does the approach and framework used to develop and implement the data warehouse project follow any industry best practices and well documented? |
| 2 | Databases | Does the database created supports huge data indexing and querying? |
| 3 | Metadata Repositories | Does the metadata repository facilitates integration and accommodate change management? |
| 4 | Design Tools | Does the tools and techniques employed to design and develop the data warehouse user friendly and ease to use? |
| 5 | Extract, Transfer and Loading | How effective and efficient the ETL tool and process are? |
| 6 | Reporting | Does the solution features any ad hoc queries or reporting tool? |

| 7 | OLAP | Does the solution provide any inline or sequential analytics that can be leveraged as part of a business process? |
|---|---|---|
| 8 | Data mining | Does the solution offers key data mining functions such as data modelling, multidimensional analysis, scoring and visualisation? |
| 9 | Data Mart Suites | Does the solution integrated with other components of a data mart? |

On the other hand, Marc Demarest suggested that the evaluation criteria for data warehousing shall begin with a look back to where the data warehouse market began, i.e. To *Bill Inmon's Building The Data Warehouse*. Demarest proposed all the operational evaluation criteria apply to an online transaction processing (OLTP) system may also apply equally to a data warehouse, including the following [6]:

- Boundary and capability of the data warehouse technology's physical storage capability.
- Loading and indexing performance of the data warehouse system.
- Operational integrity, reliability and manageability of the data warehouse system.
- Data connectivity support of the data warehouse system.
- Query processing performance of the data warehouse system.

Milicevic and Batos [7] also suggested that the basic parameters for evaluation of a data warehouse are Extract, Transform and Loading (ETL) process speed, disk space consumption, query performances and user friendliness.

In this research work, an evaluation model based on Bill Inmon's definition of data warehouse is proposed to evaluate the Cancer data warehouse.

III. THE EVALUATION MODEL

Data warehouse is defined as a "*subject-oriented, integrated, time-variant and non-volatile collection of data in support of management's decision making process*" by the father of the data warehouse, Bill Inmon [3], in which:

i. *Subject-oriented*: All relevant data concerning a particular subject are gathered and stored in a single database.

ii. *Integrated*: Data that is gathered from a variety of data sources and merged into the data warehouse must be consistent in format, naming and other aspects. They must resolve such problems as naming conflicts and inconsistencies among units of measure. When they achieve this, they are said to be integrated.

iii. *Time-variant*: Data in a data warehouse are tagged with time to support both current and historical perspective measurements. Data are stored in the data warehouse for a long period of time, to facilitate the analysis of trends and relationship between data, in an attempt to improve decision making process.

iv. *Non-volatile*: Data in a data warehouse always stay static or stable to enable a highly consistent dimensional view of data. There is no modification or deletion performed against the data after it has been loaded into the database. Due to non-volatility, data is maintained in a consistent fashion, and data warehouse can be heavily optimized for query processing.

Holding Inmon's definition closely, the developed Cancer data warehouse is examined in 2 aspects, i.e. (a) Data characteristic and (b) Operational perspectives. The evaluation criteria are explained blew:

(a) The data characteristic

- *Reliability (Subject- oriented):* Does the data evolve around a particular subject? and How reliable are the data stored in the data warehouse?
- *Integrity (Integrated):* Does the data consistent in format?
- *Capability (Time-variant):* Does the data tagged with a time?, Can the data warehouse host both historical and current data? and What is the maximum physical storage of the data warehouse?
- *Managaebilty (Non-volatile)*: How is the data maintained in the data warehouse? and Is there any tool to manage and fine tune the data warehouse?

(b) The Operational perspectives

- *Data Extraction (Subject- oriented):* How useful does the data extraction technique searches the relevant data?
- *Data cleansing (Integrated):* How effective does the developed data cleansing technique cleanses the raw data and converts them into the proper format for data mining process?
- *Data loading (Time-variant):* How helpful does the developed data loading technique captures raw data from various data sources?
- *Data querying (Non-volatile)*: How well does the developed data querying technique performs on table selection, complex SQL queries and MDX?

IV. THE CANCER DATA WAREHOUSE

This paper reviews the evaluation of healthcare data warehouse specific to cancer diseases. The cancer diseases are selected as the subject matter of the healthcare data warehouse research work.

A. *The cancer data warehouse architecture*

The Data warehouse architecture is a description of the components of the warehouse, with details showing how the components will fit together [8]. Figure 1 shows a typical architecture of a data warehouse system which includes three major areas that consist of tools for extracting data from multiple operational databases and external sources for cleaning, transforming and integrating this data; and loading data into the data warehouse[9]. Data is imported from several sources and transformed within a staging area before it is integrated and stored in the production data warehouse for further analysis.

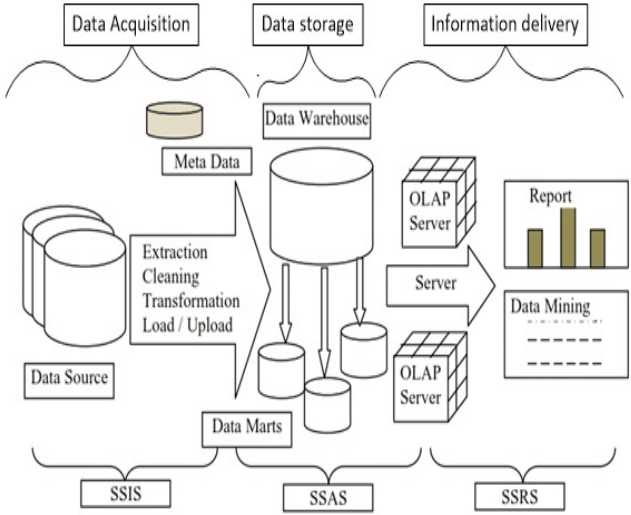

Figure 1. Data warehouse architecture

There are three major areas in the data warehouse architecture as following:

- Data acquisition.
- Data storage.
- Information delivery.

The detailed star schema in figure 2 below illustrates the data architecture of the cancer data warehouse [1].

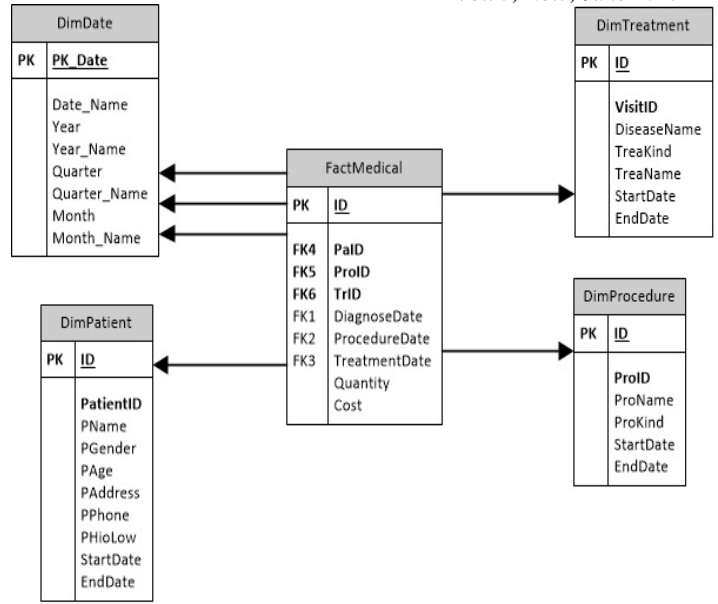

Figure 2. Star schema for cancer data warehouse

Then we will explain in details the three above areas for architecture according to the star schema:

*1) Data Acquisition area:* This covers the entire process of extracting data from the Access database, medical files such as (patient medical records, blood tests, urine test results, x-ray results and etc.) then moving all the extracted data to the staging area and preparing the extracted data for loading into data warehouse repository in this area there are a set of function and services.

*a) Data Extract-Transform-Load (ETL):* using Microsoft SQL Server integration services (SSIS) to select data and transform it to appropriate format then load this data into data warehouse [10, 11].

*b) Data Cleansing :* Before being loaded into the data warehouse, data extracted from the multiple sources was cleaned using built-in Fuzzy lookup tools contained in SSIS.

*2) Data Storage area:* This covers the process of loading the transformed data from the staging area into the data warehouse repository. Microsoft SQL Analysis Services (SSAS) used to perform this operation and convert data to multidimensional data [12, 13].

*3) Information delivery area  :* The information delivery component makes it easy for the doctors and decision making to access the information directly from the data warehouse. Microsoft SQL Reporting Services (SSRS) used to perform this operation [14, 15].

## V. THE EVALUATION RESULT

### A. Evaluation of data characteristics

The developed Cancer data warehouse is evaluated against the 4 data characteristics below:

*1) Reliability (Subject- oriented):* In this healthcare data warehouse research work, the cancer diseases are selected as the subject matter. All data that are related to the diagnosis and treatment recommendation decision making process for Cancer disease, such as the cancer types, phases, stages, treatments as well as patients' details, are stored in the data warehouse. This subject orientation promotes an easy to understand data presentation format for doctors and other health professionals.

*2) Integrity (Integrated):* To maintain a minimum level of integrity in the Cancer data warehouse, the following techniques are used:
- Integration between Access data and Excel file and then load the integrated data into Microsoft SQL Server
- Use of cross reference tables to represent multiple instances of the same information with a single instance in the database.
- Auto-generate and store primary keys in Integer, such as ID in DimPatient table, ID in DimProcedure table and etc, to avoid human error.

*3) Capabiltiy (Time-variant):* To facilitate both current and historical perspective measurements, a "startDate and EndDate" fields are used to tag the patient's medical data (i.e. the DimPatient table).

*4) Manageabilty (Non-volatile):* Data in the Cancer data warehouse are stable; they are appended to the Cancer data warehouse without deletion. There is no restriction on the insertion of data, and all primary keys are automatically created using the IDENTITY(1,1) value provided by the MSQL database.

### B. Evaluation of data operations

The developed data warehousing techniques are assessed from the 4 data warehouse operational perspectives below:

*1) Data extraction performance (Subject- oriented):* Data are manually extracted from Access data base and excel files. This manual data extraction approach is time consuming. A lot of effort is given to identify relevant, vital and valuable information as well as to filter out incomplete or erroneous information within the mass data from sources. The SSIS solve this problem.

*2) Data Cleansing performance (Integrated):* SSIS is used as a tool to cleanse and transform the extracted raw data to ensure data quality and integrity, before they are loaded onto the data warehouse. Before being loaded into the data warehouse, data extracted from the multiple sources was cleaned using built-in Fuzzy lookup tools contained in SSIS. The fuzzy look up is used to lookup data rows with spelling mistakes and correct such mistakes also delete duplicated data. The SSIS package is developed to cleanse the medical data and transform them to the appropriate format.

*3) Data Loading performance (Time-variant):* Medical data are loading into data warehouse after transform and cleanse this data to formats that predefined for data warehouse.

*4) Data Querying performance (Non-volatile):* The data querying performance should be evaluated from the aspects below:

*a) Data response time:* Data querying response time is often depending on the (i) size of data, (ii) complexity of query and (iii) specification of hardware. The larger the data processed or the more complex the queries, the slower the response time. In our Cancer data warehouse project, a satisfactory query response time is obtained as approximate three second for a complex query to retrieve data from data warehouse resided on a machine with single processor and 3 GB memory.

The Figure 3 show the response time from executing a complex query in data warehouse and database to show the same result.

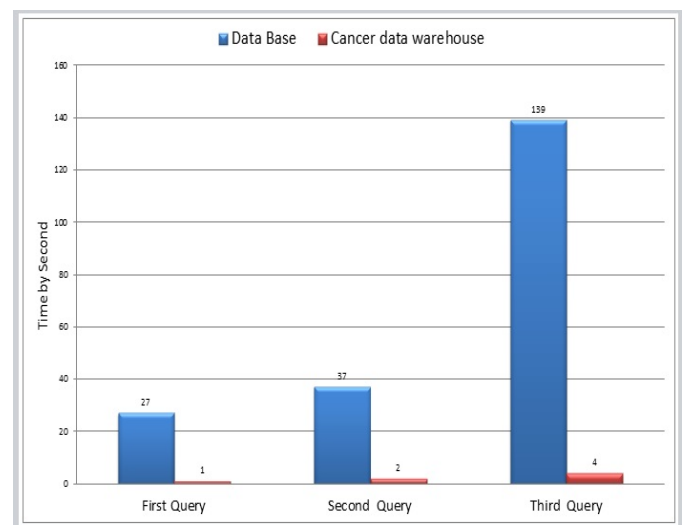

Figure 3. A response time comparsion between cancer data warehouse and data base

*b) Data visualisation:* By using MDX query language [16] as the data query and presentation tool, the visual interpretation of complex relationships of the multidimensional data can only be presented in a tabular format. However, the data query techniques and drill down capability only facilitate advanced users who have basic SQL knowledge. To extend the data query facilities to other users and to enhance data visualization advanced data presentment and graphics tools should be considered to illustrate data relationships and provide a graphical presentation of information in histograms, pie charts and bar graphs, maps etc.

## VI. CONCLUSION

The healthcare industry is one of the world's largest, fastest-developing and most information-rich industries for take advantage from this information we build the Cancer data warehouse to integrate between the operational data base and medical files and therefore the analysis on data makes easy by using OLAP cubes and viewing multilevel of details from the data. Then we can analyses the cancer diseases, the cost of treatment for these diseases, Death rate in specific type of cancer and the impact of a particular drug on the disease. The cancer data warehouse built with Microsoft technology (Microsoft Sql Server 2008 and C# language).

The cancer data warehouse evaluation is examined according to Bill Inmon's definition as "*data warehouse subject-oriented, integrated, time-variant and non-volatile collection of data in support of management's decision making process*". All conditions in this definition have been achieved.

AUTHORS PROFILE

Dr. Osama E. Sheta; PhD (Irkutsk State Technical University, Irkutsk, Russia).
Assistant Professor (IS) Faculty of Science,
Zagazig University, Egypt.
Assistant Professor (IS) Faculty of Science,
Al Baha University, KSA.

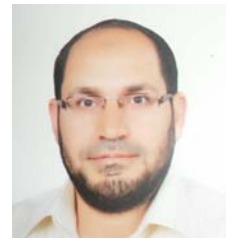

Mr. Ahmed Nour Eldeen; BS (Mathematics and
Computer Science); University of Zagazig,
Egypt (2006) and pre-MSc titles in computer
science from Zagazig University ( 2011).
Director of information systems
at Health Insurance Organization Egypt
(Elsharkiya branch) and has a strong
background in Relational Database Systems,
Database Administration ASP.NET and
C# language.

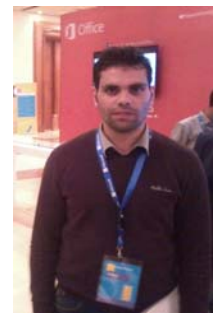